\documentclass[aps,final,superscriptaddress,twocolumn]{revtex4-2}

\usepackage{amsmath,amssymb}
\usepackage{url}
\usepackage{graphicx}
\usepackage{hyperref}
\usepackage{soul}
\usepackage{gensymb}
\hypersetup{
  unicode = false,
  colorlinks= true,
  urlcolor= black,
  linkcolor= blue,
  citecolor= blue,
  pdftitle= {CREASE2D Manuscript}
}
	
\usepackage{xcolor}
\setcitestyle{super}

\makeatletter
\def\@fnsymbol#1{\ensuremath{\ifcase#1\or \dagger\or \ddagger\or \mathsection\or \mathparagraph\or \|\or **\or \dagger\dagger \or \ddagger\ddagger \else\@ctrerr\fi}}
\renewcommand\frontmatter@abstractwidth{\dimexpr\textwidth\relax}
\makeatother

\begin{document}
\title{Computational Reverse Engineering Analysis of Scattering Experiments Method for Interpretation of 2D Small-Angle Scattering Profiles (CREASE-2D)}
\author{Sri Vishnuvardhan Reddy Akepati}
\thanks{These authors contributed equally to this work.}
\affiliation{Data Science Program, University of Delaware, Newark, DE 19716 USA.}
\author{Nitant Gupta}
\thanks{These authors contributed equally to this work.}
\affiliation{Department of Chemical and Biomolecular Engineering, University of Delaware, Newark, DE 19716 USA.}
\author{Arthi Jayaraman}
\thanks{Corresponding author. Email: arthij@udel.edu}
\affiliation{Department of Chemical and Biomolecular Engineering, University of Delaware, Newark, DE 19716 USA.}
\affiliation{Department of Materials Science and Engineering, University of Delaware, Newark, DE 19716 USA.}

\begin{abstract}
Characterization of structural diversity within soft materials is key for engineering new materials for various applications. Small-angle scattering (SAS) is a widely used characterization technique that provides structural information in soft materials at varying length scales (nm to microns). The typical output of a SAS measurement is the scattered intensity $I(\mathbf{q})$ as a function of $\mathbf{q}$, the scattered wavevector with respect to the incident wave. The scattered wavevector $\mathbf{q}$ is represented by its  magnitude $|\mathbf{q}| \equiv q$ (in inverse distance units) and  azimuthal angle $\theta$. While structures with isotropic spatial arrangement can be interpreted by analysis of azimuthally averaged one-dimensional (1D) scattering profile, to understand anisotropic spatial arrangements, one has to interpret the two-dimensional (2D) scattering profile, $I(q, \theta)$. Manual interpretation of such 2D profiles is complicated, and usually involves fitting of approximate analytical models to azimuthally averaged sections of the 2D profile. In this paper, we present a new method called CREASE-2D that interprets, without any azimuthal averaging, the entire 2D scattering profile, $I(q, \theta)$, and outputs the relevant structural features. CREASE-2D is an extension of the `computational reverse engineering analysis for scatting experiments' (CREASE) method that has been used successfully to analyze 1D SAS profiles for a variety of soft materials. CREASE uses a genetic algorithm for optimization and an artificial neural network (ANN) as the surrogate machine learning (ML) model for fast calculation of 1D `computed' scattering profiles that are then compared to the experimental 1D scattering profiles in the optimization. CREASE-2D goes beyond CREASE by enabling analysis of 2D scattering profiles, which is far more challenging to interpret than the azimuthally averaged 1D profiles. Further, we use XGBoost as the surrogate ML model in CREASE-2D, in place of ANNs, to relate structural features to the $I(q, \theta)$ profile. The CREASE-2D workflow identifies the structural features whose computed $I(q, \theta)$ profiles, calculated using the surrogate ML model, match the input experimental $I(q, \theta)$. We test the performance of CREASE-2D by using as input a variety of \textit{in silico} 2D scattering profiles whose structural features are known to us.  We demonstrate that CREASE-2D works well by showing that for every one of these input \textit{in silico }2D scattering profiles, CREASE-2D converges towards the correct structural features. We expect this CREASE-2D method will be a valuable tool for materials' researchers who need direct interpretation of the 2D scattering profiles in contrast to analyzing azimuthally averaged 1D $I(q)$ vs. $q$ profiles that can lose important information related to structural anisotropy.
\end{abstract}

\maketitle

Researchers studying soft materials, namely polymers, colloids, liquid crystals, gels and chemical formulations, aim to establish molecular design-structure-property relationships to engineer new materials with improved physical properties. Towards this goal, microscopy and scattering are two prominent characterization techniques for understanding the structure formed within such soft materials. Microscopy techniques that are commonly used for soft materials include optical microscopy, to probe structures with length scales above 10 microns, and scanning-electron/transmission-electron/atomic-force microscopy (SEM/TEM/AFM) to probe structures with features below 10 microns. Such microscopy methods can reveal the pertinent structural features in the area of the material that is imaged, albeit only over a narrow range of length scales. Furthermore, microscopy only outputs a two-dimensional (2D) projection of the structure and the depth information can be non-trivial to interpret. In contrast, bulk structural characterization techniques that rely on scattering of light (X-rays/visible/infrared) or neutrons are able to reveal three dimensional (3D) structural information across multiple length scales. In particular, for soft materials, small angle X-ray scattering (SAXS) and small angle neutron scattering (SANS) techniques  \cite{wei2021characterizing,grawert2020structural,jeffries2021small,lombardo2020structural,semeraro2021increasing,bressler2015sasfit,doucet2017sasview,pedersen1997analysis} are used widely to elucidate spatial distributions of amorphous (i.e., not crystalline) ordered or disordered structures at various length scales. 

A typical SAXS or SANS measurement captures the scattered intensity $I(\mathbf{q})$ as a function of the scattered wavevector $\mathbf{q}$ with respect to the incident wave, expressed by its magnitude $|\mathbf{q}| \equiv q$ (in inverse distance units) and  the azimuthal angle $\theta$. For materials that have isotropic structural arrangements, the patterns found in the 2D SAXS/SANS profiles, $I(q,\theta)$, are expected to exhibit spherical or cylindrical symmetry. Analysis of such symmetric scattering profiles involves integrating over all azimuthal angles and fitting analytical models to the one-dimensional (1D) form of the scattering profile - $I(q)$ vs. $q$. Presence of a peak in these 1D scattering profiles at a certain $q$ value indicates presence of structural correlations at length scales around $2\pi/q$, either due to the dimensions of the constituent particles (i.e., form factor, $P(q)$) or the arrangement of particles that influences their inter-particle spacing (i.e., structure factor, $S(q)$). Therefore, even when a structure consists of anisotropic particles that are devoid of any inter-particle orientational order, the 1D scattering profile can reveal most of the relevant structural details about the material. Further, in case of dilute solutions (e.g., amphiphilic polymer solutions at low polymer concentrations), the form factor of the primary particle (e.g., assembled micelles) can be analyzed using shape-dependent or shape-independent models to obtain the dimensions of the primary particles.\cite{POKORSKI2019157} However, in cases where there is significant dispersity in dimensions of the primary particles, such models can be poor approximations and the resulting analysis can be flawed. In case of concentrated solutions (e.g., amphiphilic polymer solutions at high concentrations), in addition to the form factor of the primary particle, one has to analyze the structure factor which holds the information about spatial arrangement of the primary particles (e.g., inter-particle or inter-micelle arrangement). In such cases, if one assumes that the form factor remains constant with changing concentration, then one can use the analyzed form factor obtained at dilute concentration to interpret the structure factor.  Analytical structure factor models like the `sticky hard sphere' or `Percus Yevick' \cite{hammouda2008probing} can be used to interpret isotropic structures with low dispersity. However, in the case of systems where the values of primary particle's shape and size or entire distributions of shape and size of primary particles change with concentration or systems in which the structure develops anisotropy during processing or rheological measurements\cite{eberle2012flow,gordon2021structural,richards2017strain}, the interpretation of these scattering profiles can be challenging.  

To circumvent these challenges with traditional approaches involving manual fitting with shape-dependent or independent models that can be approximate or incorrect in some cases, there is a need for other analysis approaches. Additionally, the surge in high-throughput measurements and the quest for artificial intelligence (AI) driven manufacturing demand analyses methods that can be fast and automated in interpreting scattering profiles, and complementary characterization results, as and when the measurement is done. We direct readers to a recent perspective by Anker \textit{et al. }that covers many ongoing developments and studies within this topic of fast computational analysis of scattering and spectroscopic measurements in materials sciences. \cite{Anker2023ChemSci}. The challenges for computational methods being developed for fast and/or automated scattering analyses in the area of synthetic soft materials are different from inorganic hard materials or biological molecules. This is because (non-biological) soft materials structures tend to be mostly amorphous, often exhibiting significant dispersity in structural dimensions, unlike the precise crystalline order seen in inorganic materials or secondary and tertiary structures of proteins,\cite{friesner2013computational,dorn2014three,Franke2018BiophsyJ} . To address this specific need in the area of soft materials with amorphous structures, Jayaraman and coworkers recently developed the `Computational Reverse Engineering Analysis of Scattering Experiments' (CREASE) method.\cite{beltran2019computational,heil2023computational,heil2022computational,wessels2021computational,wessels2021machine,wu2022machine,ye2021computational,heil2021computational,crease_ga} 

The CREASE method outputs the features or descriptors of the three-dimensional (3D) structures that produce `computed' scattering profiles $I_\text{comp}(q)$ which closely resemble the scattering profile obtained in experiments $I_\text{exp}(q)$. Rather than iterating exhaustively over 3D structures themselves, a computationally intensive and slow process, in CREASE the optimization cycle iterates over a lower dimensional representation of the 3D structures. We call these lower-dimensional descriptors of structure as \emph{structural features}; as we use genetic algorithm for optimization, in the jargon of evolutionary algorithms, we  refer to these structural features  as``genes''.

In a typical GA optimization loop, an initial population of ``individuals'' is generated, where each individual has a unique set of structural features or ``genes''. The structural features can have single values of structural parameters or encode parameters representing distributions of structural parameters. For each individual, these structural features are converted to a computed scattering profile using a surrogate machine learning (ML) model. The computed scattering profile $I_\text{comp}(q)$ of each individual is then compared to the input scattering profile $I_\text{exp}(q)$. The extent of match between the $I_\text{comp}(q)$ and $I_\text{exp}(q)$ is calculated as a \textit{fitness} value for that individual. After the fitness value has been calculated for all individuals in a generation, then a new ``generation'' of individuals is created based on the current generation's fitness ranking and genetic operations like``pairing'' and ``mutations''\cite{burke2004diversity}. As the optimization proceeds, with each new``generation'', the individuals progressively exhibit better fitness values, i.e., improvement in the match between their $I_\text{comp}(q)$ and the input $I_\text{exp}(q)$. At the end of the GA cycle, upon convergence in fitness values, CREASE outputs multiple individuals (i.e., sets of structural features) that all have the mutually similar scattering profiles that also match the input scattering profile. If the GA results consist of  multiple distinct sets of structural features, then one would either use their domain knowledge or guidance for imaging techniques and/or molecular modeling and simulations to remove the ``individual(s)'' that are deemed unphysical and keep only those ``individual(s)" that are physically possible. 

Within the optimization loop, the use of surrogate ML models for calculation of $I_\text{comp}(q)$ for each individual has significantly accelerated the computational speed of CREASE. Traditionally, for 3D structures with known positional coordinates of each particle or constituents of each particle, one would use the computationally intensive Debye scattering equation to calculate scattering profiles. To accelerate this step of calculating $I_\text{comp}(q)$, in recent CREASE studies, Jayaraman and coworkers introduced the idea of using a surrogate ML model (e.g., artificial neural networks or ANN) that connects the structural features (i.e., lower dimensional representation of the 3D structure) to its $I_\text{comp}(q)$.\cite{wessels2021machine,wu2022machine,heil2022computational} Using this machine learning enhanced CREASE (ML-CREASE) one can interpret input scattering profiles fast and on modest computational resources as described in Refs. \cite{heil2023computational,heil2022computational,wu2022machine}

\begin{figure*}[t!]
    \includegraphics[width=\textwidth]{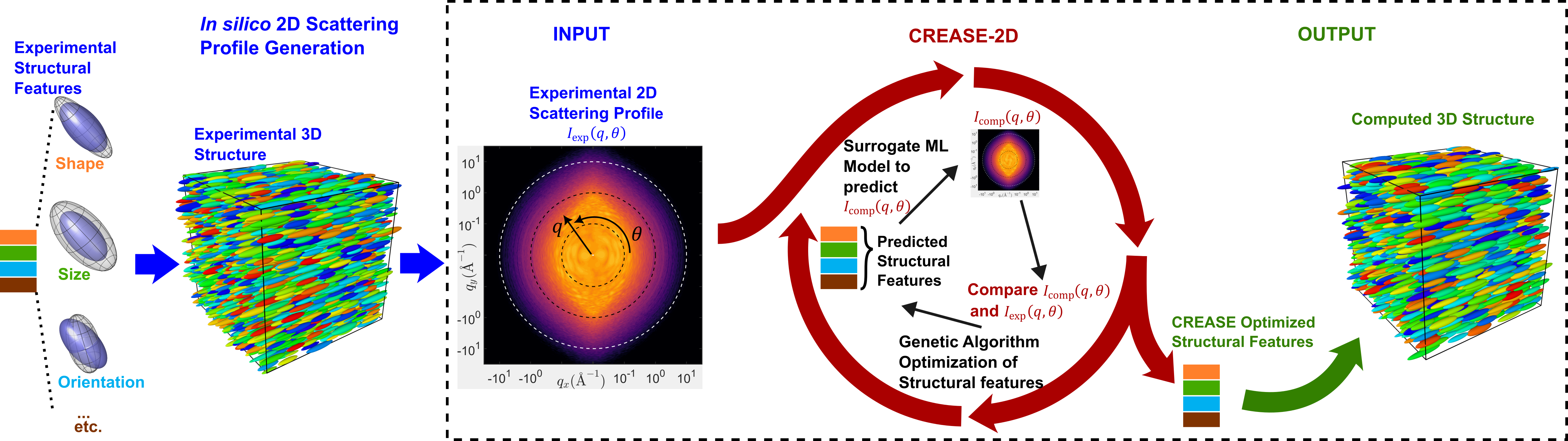}
    \caption{CREASE-2D method workflow used in this paper. For proving the CREASE-2D works correctly, we used as input an \textit{in-silico} 2D scattering profile generated from a 3D structure with predefined set of structural features. Only the 2D scattering profile, which we call ``experimental'' scattering profile or $I_\text{exp}(q,\theta)$ is used as the input to the CREASE-2D method. The genetic algorithm (GA) optimizes towards structural features whose $I_\text{comp}(q,\theta)$ closely resembles $I_\text{exp}(q,\theta)$. By comparing the optimized structural features to the ones used to create the $I_\text{exp}(q,\theta)$ we can show that CREASE-2D works well. }
    \label{fig:CREASE2D}
\end{figure*}

There have been multiple soft materials systems where CREASE has been used successfully to analyze the 1D scattering profiles, and in many cases, CREASE has performed better than existing analytical models. For example, CREASE has been used to analyze the form factor of assembled structures in dilute solutions; for example, spherical micelles\cite{beltran2019computational}, cylindrical micelles\cite{wessels2021computational,wessels2021machine} and vesicles\cite{ye2021computational} formed by novel polymers or macromolecules in solution. In these cases, the existing analytical models were either too approximate\cite{beltran2019computational} or could not handle dispersity in structural dimensions well\cite{ye2021computational}. In some cases, CREASE was used to test hypotheses of potential assembled structures in the solution which could not have been done with analytical models alone. \cite{lee2020hierarchical} CREASE has also been used to interpret the scattering profiles of concentrated particle systems where the form of the particle was known \textit{a priori} (e.g., simple spherical particles); in this case, CREASE was used to understand the extent of mixing within binary nanoparticle mixtures \cite{heil2021computational,heil2022computational}.  CREASE has also been extended to the `P(q) and S(q) CREASE' version that can analyze both the form and the structure factors of the primary particles simultaneously\cite{heil2023computational}. This `P(q) and S(q) CREASE' method was used to understand a system of silica particles coated with surfactant (core-shell particles) at varying temperature and salt concentrations. Both temperature and salts affect the cationic surfactants in the shell around the particles and as a result, the form of the surfactant-coated particles and inter-particle structure. \cite{bhuvneshLangmuir2023}

While all of the above applications of CREASE involved isotropic structures and the 1D scattering profile from experiments as input, in this paper we have extended the CREASE method to CREASE-2D that can analyze 2D scattering profiles directly and in turn, enable interpretation of structures that may have anisotropy.
 
The input to CREASE-2D is a 2D scattering profile coming from structures that have some orientational order within the material's structure either produced by processing conditions (e.g., shear or field-induced alignment of domains) and/or because of the form of primary particle (e.g., ellipsoidal domain). In such cases, characterization of structural anisotropy requires the use of the 2D SAS profiles $I_\text{exp}(q,\theta)$ which hold information of the length scales of arrangements that can vary along various azimuthal angles, $\theta$. 

In this paper, we present all the relevant details of CREASE-2D method development and demonstrate its successful application by correctly outputting the structural features of the the 3D structures that gave rise to the input \textit{in-silico} ($I_\text{exp}(q,\theta)$). 

\subsection*{ CREASE-2D: Overview and Development}
\textbf{Figure \ref{fig:CREASE2D}} provides an overview of the CREASE-2D workflow presented in this paper. The overall development of CREASE-2D involves four key steps: 
\begin{enumerate}
    \item Generating a dataset of 3D structures having an extensive variation of all important structural features. The structural features that we demonstrate in this study are distributions of domain sizes, shapes, orientational order, and volume fraction of domains that produce the scattering; 
    \item Computing the 2D scattering profiles for each of those 3D structures; 
    \item Using the combined dataset of structural features and their computed 2D scattering to train the surrogate ML model that will output a computed scattering profile for an input of structural features; and
    \item Incorporating the trained ML model within the GA optimization loop to fulfill the CREASE-2D workflow. 
\end{enumerate}

While step (4) above enables a smooth and fast execution of the CREASE-2D method, steps (1) - (3) are necessary for an accurate and reliable representation of experimentally relevant structural configurations and their scattering profiles. The amount and quality of data generated in steps (1) and (2) will also determine the accuracy of the surrogate ML model in step (3), which in turn dictates the efficacy of the CREASE-2D optimization. Before we describe each of the above steps in more detail, we note the similarities and differences between CREASE-2D and prior implementations of CREASE.

Similar to previous implementations, CREASE-2D method also uses a GA to optimize structural features. The GA loop proceeds in a similar manner as in previous uses of CREASE and stops when fitness of the individuals converges, i.e., an individual's computed 2D scattering profile $I_\text{comp}(q,\theta)$ matches the input profile $I_\text{exp}(q,\theta)$. One difference between the previous CREASE implementation and CREASE-2D is in the choice of the surrogate ML model to calculate the $I_\text{comp}(q,\theta)$ for each individual. The surrogate ML model needed in this case not only needs to handle input in the form of a table having multidimensional variation of its structural features, but also output a 2D scattering profile rather than a 1D scattering curve of $I_\text{comp}(q)$ vs $q$. More details are provided in the steps (3) and (4) sub-sections below.

\subsection*{Step (1): Generating a dataset of 3D structures with varying structural features}

To develop a reliable surrogate ML model for linking  structural features to the 2D scattering profile, we need a training dataset that contains sufficient samples of 3D structures with all potential variations in structural features that influence their computed 2D scattering profiles. This sub-section describes this process of generating such a dataset. 

In principle, structural features condense the detailed representation (e.g., x, y, and z coordinates of all particles) of the 3D structure to a few numerical values that pertain to the distributions of parameters describing the 3D structure. In ML jargon, structural features are similar to the lower dimensional latent space variables encoding a higher dimensional input function. Our philosophy is that the structural features should be information that a soft materials researcher would understand and find relevant. By relevance we mean that the interested structural features would be ones (e.g.,  shapes, sizes, and spatial arrangement of the domains, extent of mixing/demixing within domains/between domains, orientational alignment of domains, grain boundaries, etc.) that will likely control properties/function of the soft material. Thus, we choose not to have automatically encoded latent space variables that lack a physical meaning and not easily interpreted by human, and instead define our own structural features using our soft materials domain knowledge.

To demonstrate our choices of structural features for a representative example of soft materials with structural anisotropy, we consider a model system of spheroidal particles with well-defined distributions of shapes, sizes, and orientations (shown schematically in \textbf{Figure \ref{fig:CREASE2D}}), along with variations in the particles' packing fractions in the material. Generation of such 3D structures is facilitated by our recently developed (open-source) computational method called CASGAP (Computational Approach for Structure Generation of Anisotropic Particles)\cite{gupta2023}. In the original manuscript \cite{gupta2023}, we  demonstrated the versatility of the CASGAP method to generate 3D structures for user-provided distribution of particle sizes, shapes, and orientations at or close to the target volume fraction.  Accordingly, the CASGAP method uses parameters $R_\mu$, $R_\sigma$, $\gamma_\mu$, $\gamma_\sigma$, and $\kappa$ to generate the 3D anisotropic structure with a target $\phi_\text{target}$. These structural descriptors serve as the structural features for use in this development of the CREASE-2D workflow. While the detailed description of these structural features can be found in the original manuscript \cite{gupta2023}, we review some relevant details below:
\begin{enumerate}
        \item The particle sizes and shapes are expressed by the spheroidal volumetric radius $R = \sqrt[3]{a^2c}$ and the spheroidal aspect ratio $\gamma=c/a$, where $a$ and $c$ are the lengths of the semi-minor and semi-major axes of the spheroid, respectively. As done in the original manuscript\cite{gupta2023}, the variations in size and shape are modeled by a log-normal distribution, each with their means ($R_\mu$, $\gamma_\mu$) and standard deviations ($R_\sigma$, $\gamma_\sigma$). These quantities provide us with the first four structural features for CREASE-2D.
        \item The orientations in the structure is quantified by a 3D vector pointing along the major axis $\mathbf{V}$ of the spheroid. With such description of orientations, we adopt the 3D von Mises-Fisher (vMF) distribution (see details in Ref. \cite{gupta2023}) to model the distribution of orientational order expressed succinctly by the $\kappa$ parameter. The $\kappa$ parameter is a measure of the inverse-dispersity in orientation and is defined around a preferred orientation $\mathbf{\Lambda}$. $\kappa=0$ indicates complete lack of orientational order (i.e., $\mathbf{V}$ is uniformly distributed on the surface of a sphere) and $\kappa\rightarrow\infty$ indicates perfect orientational order (i.e., $\mathbf{V}=\mathbf{\Lambda}$). Relying on the premise that for an anisotropic structure, the principal axes of anisotropy can be aligned with the laboratory frame of reference during scattering measurements such that $\mathbf{\Lambda}=\mathbf{\hat{x}}$, enables us to use only $\kappa$ as the fifth structural feature.
        \item Lastly, the concentration of particles is quantified by the volume fraction of particles, $\phi$. If dense particle configurations is desired, a trade-off is observed between the computational time for structure generation and the value of volume fraction achieved in that time. The CASGAP method is designed with this trade-off in mind and can be terminated at any point of the structure generation while maintaining a structure that adheres to the desired structure features' distribution. However, in such cases of early termination, the actual volume fraction $\phi$ may not reach the value of $\phi_\text{target}$ leading to $\phi\le\phi_\text{target}$. With such an expectation, we use the actual $\phi$ evaluated after the structure is generated as the sixth structural feature since the scattering profile computed in step (2) (described in the next sub-section) can be significantly influenced by the actual volume fraction of the particles.
\end{enumerate}  

\begin{figure*}[t!]
    \centering
    \includegraphics[width=1\textwidth]{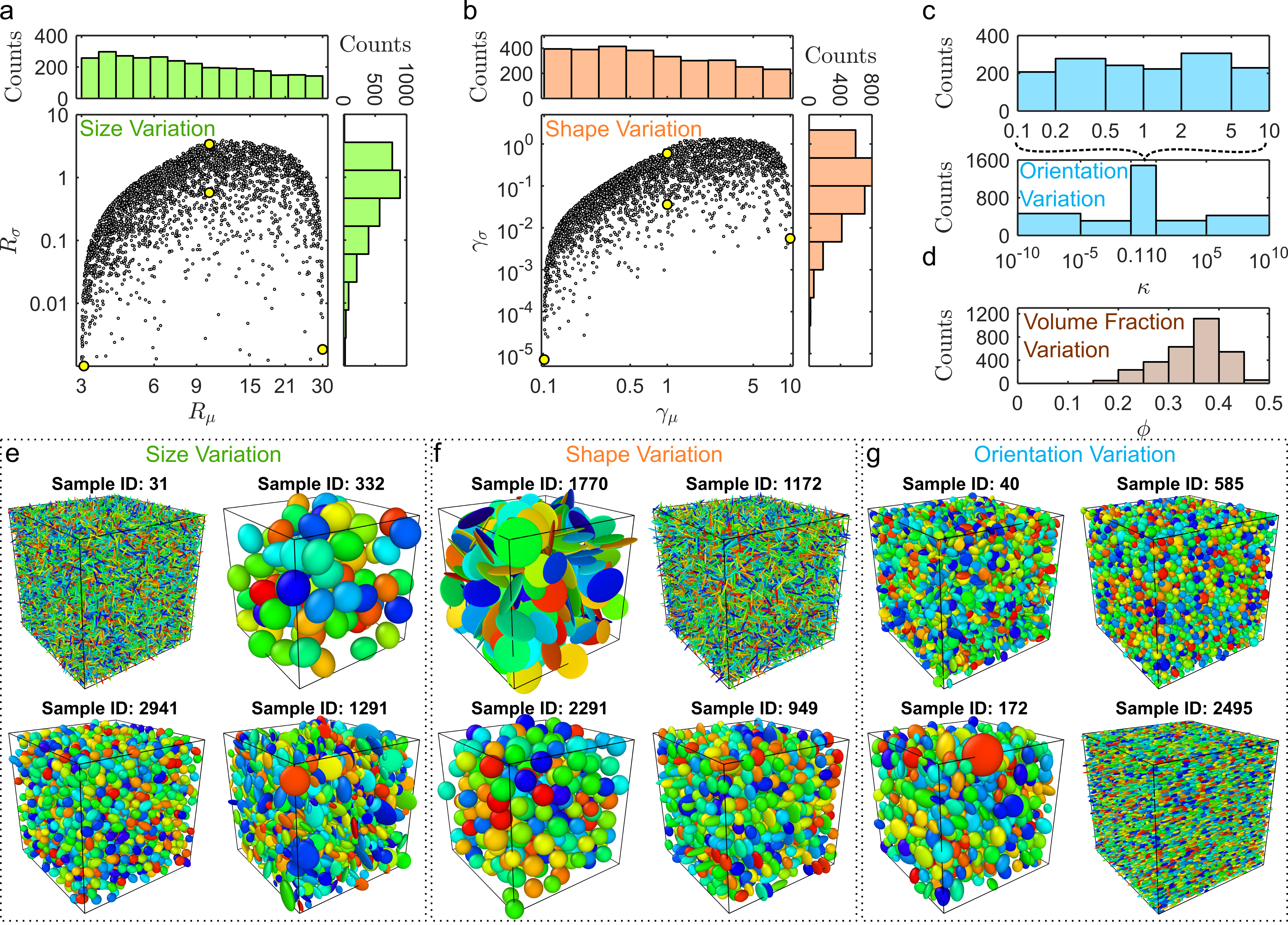}
    \caption{(a) Histograms and scatter plots describing how the mean and standard deviations of the volumetric radius $R$ is varied in each of the 3000 samples in the dataset. (b) Similar to (a) but for the mean and standard deviations of the aspect ratio $\gamma$. (c) Histograms describing the distribution in the orientational anisotropy parameter $\kappa$ for the von-Mises Fisher distribution\cite{gupta2023}. The histogram in the range of $0.1 \le \kappa \le 10$ is shown separately (on top) to indicate the distribution of nearly $50\%$ samples uniformly drawn from this range. (d) Histogram describing the distribution in the volume fraction $\phi$ of the generated structures. (e-g) Representative snapshots of 3D structures drawn from the dataset, showing size, shape, and orientation variations, respectively. Use of different colors facilitate easy distinction of individual particles visually. The structures in (e) and (f), correspond to points highlighted in the scatter plots of (a) and (b), respectively. The detailed information about their structural features are provided in Table~\ref{table:strucfeaturesfig2}.}
    \label{fig:samplingstrucfeatures}
\end{figure*}

\begin{table}[ht]
\centering
\caption{\textbf{Values of Structural Features For Select Few Samples (in the same order as) shown in Figure~\ref{fig:samplingstrucfeatures}.}}
\begin{tabular}[t]{crrrrrr}
\hline
\textbf{Sample ID} &$R_\mu$ &$R_\sigma$ &$\gamma_\mu$ &$\gamma_\sigma$ &$\kappa$ &$\phi$\\
\hline
\textbf{31}&\textbf{3.07}&0.00101&7.48&0.263&$3.11\times 10^{-1}$&0.245\\
\textbf{332}&\textbf{30.0}&0.00185&0.785&0.0352&$4.41\times 10^{-1}$&0.389\\
\textbf{2941}&10.2&\textbf{0.576}&1.28&0.200&$5.89\times 10^{-10}$&0.361\\
\textbf{1291}&10.2&\textbf{3.38}&0.488&0.176&$6.31\times 10^{0}$&0.401\\
\textbf{1770}&17.4&1.87&\textbf{0.100}&0.000&$2.18\times 10^{-2}$&0.237\\
\textbf{1172}&3.57&0.201&\textbf{9.96}&0.00558&$1.36\times 10^{-1}$&0.194\\
\textbf{2291}&19.5&2.19&1.00&\textbf{0.0358}&$1.35\times 10^{8}$&0.422\\
\textbf{949}&12.9&0.802&1.00&\textbf{0.591}&$3.83\times 10^{4}$&0.409\\
\textbf{40}&7.80&1.50&0.511&0.0696&$\mathbf{1.09\times 10^{-10}}$&0.430\\
\textbf{585}&6.80&1.63&0.881&0.0669&$\mathbf{1.23\times 10^{-1}}$&0.461\\
\textbf{172}&14.1&3.69&0.654&0.122&$\mathbf{9.70\times 10^{0}}$&0.459\\
\textbf{2495}&3.64&0.0662&2.99&1.24&$\mathbf{9.68\times 10^{9}}$&0.357\\
\hline
\end{tabular}
\footnotesize{Bolded text is used to highlight the relevant structural features depicted in Figure~\ref{fig:samplingstrucfeatures}.}
\label{table:strucfeaturesfig2}
\end{table}%

Leveraging the computational efficiency of the CASGAP method, we generate a dataset of 3000 three-dimensional structures. This dataset has a numerical index from $1-3000$  used as their Sample ID along with numerical values of all their structural features. We share some examples from this dataset openly on Zenodo.\cite{Zenodo} In \textbf{Figure~\ref{fig:samplingstrucfeatures}} we describe how each of the structural features are varied. Since each structural feature represents a physically relevant quantity with significant influence over the morphology of the particles, these could not simply be varied using a uniform distribution over their respective ranges. As a result, some of these quantities have a normal-like or a skewed distribution in their chosen ranges as shown in the plots in \textbf{Figure~\ref{fig:samplingstrucfeatures}a-d}. The numerical details of the random sampling, which is a version of Monte Carlo sampling, is discussed in detail in the \textbf{Supporting Information Section S1}. We represent all our structures by a cubic representative volume of length $L=300$~distance units (in this study 1 distance unit corresponds to 1 \AA  , but this correspondence can be changed to a different length-scale, as desired). In \textbf{Figure~\ref{fig:samplingstrucfeatures}e-g} and the accompanying \textbf{Table~\ref{table:strucfeaturesfig2}} provide some representative structure snapshots along with their structural features. Some extreme values of structural features have been indicated in Table~\ref{table:strucfeaturesfig2} with bold font; we selected the Sample IDs with these extreme values of structural features to visualize their effects on the overall structure.

The mean volumetric radius $R_\mu$ is nearly uniformly sampled over a range of $3$~\AA~to $30$~\AA, representing a variation of $1\%~L$ to $10\%~L$ (as shown in the histogram of \textbf{Figure~\ref{fig:samplingstrucfeatures}a}). To keep the size variation reasonable within the prescribed log-normal distributions, the standard deviation of volumetric radius is controlled by the mean value, such that whenever $R_\mu$ approaches its extreme values, i.e., $3$~\AA~or $30$~\AA,  $R_\sigma\rightarrow0$. This is shown in the scatter plot of \textbf{Figure~\ref{fig:samplingstrucfeatures}a}, where an envelop shape over the $R_\sigma$ distribution is observed. In \textbf{Figure~\ref{fig:samplingstrucfeatures}e} Sample 31 and Sample 332 depict the structure when $R_\mu$'s are $\sim3$ and $\sim30$, respectively. While Sample 2941 and Sample 1291 (with similar $R_\mu$) depict the extreme values of $R_\sigma$.

\textbf{Figure~\ref{fig:samplingstrucfeatures}b} shows the variation in aspect ratios in range of $1/10$ to $10$. Since this is a ratio, the values below 1 (representing oblate spheroids) are analogous, by a reciprocal relationship, to those above 1 (representing prolate spheroids). To ensure fair sampling of both these shape types, the values are nearly uniformly sampled over the logarithmic scale between $1/10$ to $10$ as shown in the histogram of \textbf{Figure~\ref{fig:samplingstrucfeatures}b}. Here, too, we ensure that whenever the mean aspect ratio $\gamma_\mu$ approaches the extremes, $\gamma_\sigma$ approaches 0 as seen from the scatter plot in \textbf{Figure~\ref{fig:samplingstrucfeatures}b}. In \textbf{Figure~\ref{fig:samplingstrucfeatures}e} Sample 1770 and Sample 1172 depict the structure when $\gamma_\mu$'s are $\sim0.1$ and $\sim10$, respectively. While Sample 2291 and Sample 949 (with similar $\gamma_\mu$) depict the extreme values of $\gamma_\sigma$.

To vary the degree of orientational order, the $\kappa$-parameter (\textbf{Figure~\ref{fig:samplingstrucfeatures}c}) can be varied by sampling equally from 4 intervals defined by the end points: $10^{-10}$, $0.1$, $1$, $10$ and $10^{10}$. Here values $10^{-10}\approx 0$ and $10^{10}\approx\infty$, are chosen to sample the perfectly isotropic and anisotropic structures, respectively. Structures from each of these intervals are in \textbf{Figure~\ref{fig:samplingstrucfeatures}e} with Sample IDs 40, 585, 172 and 2495, where $\kappa$ values are nearly $10^{-10}$, $0.1$, $10$ and $10^{10}$, respectively.

In \textbf{Figure~\ref{fig:samplingstrucfeatures}d} the  histogram shows the variation in $\phi$ for the entire dataset. Unlike all other structural features, the distribution of $\phi$ is not prescribed but is a result of CASGAP structure generation with $\phi_\text{target}=0.5$ as explained previously. If a stricter control over $\phi$ is desired, more samples at lower $\phi$ can easily be generated and added to the dataset to change the shape of the distribution.

Having the dataset of 3D structures, we calculate each of their 2D scattering profiles in Step (2). The 2D scattering profiles and the structural features then become the intended ``output'' and ``input'' data for training and testing the surrogate ML model in Step (3).

\subsection*{Step (2): Calculating 2D Scattering Profile for Each 3D Structure}

\begin{figure}[t!]
    \centering
    \includegraphics[width=0.91\columnwidth]{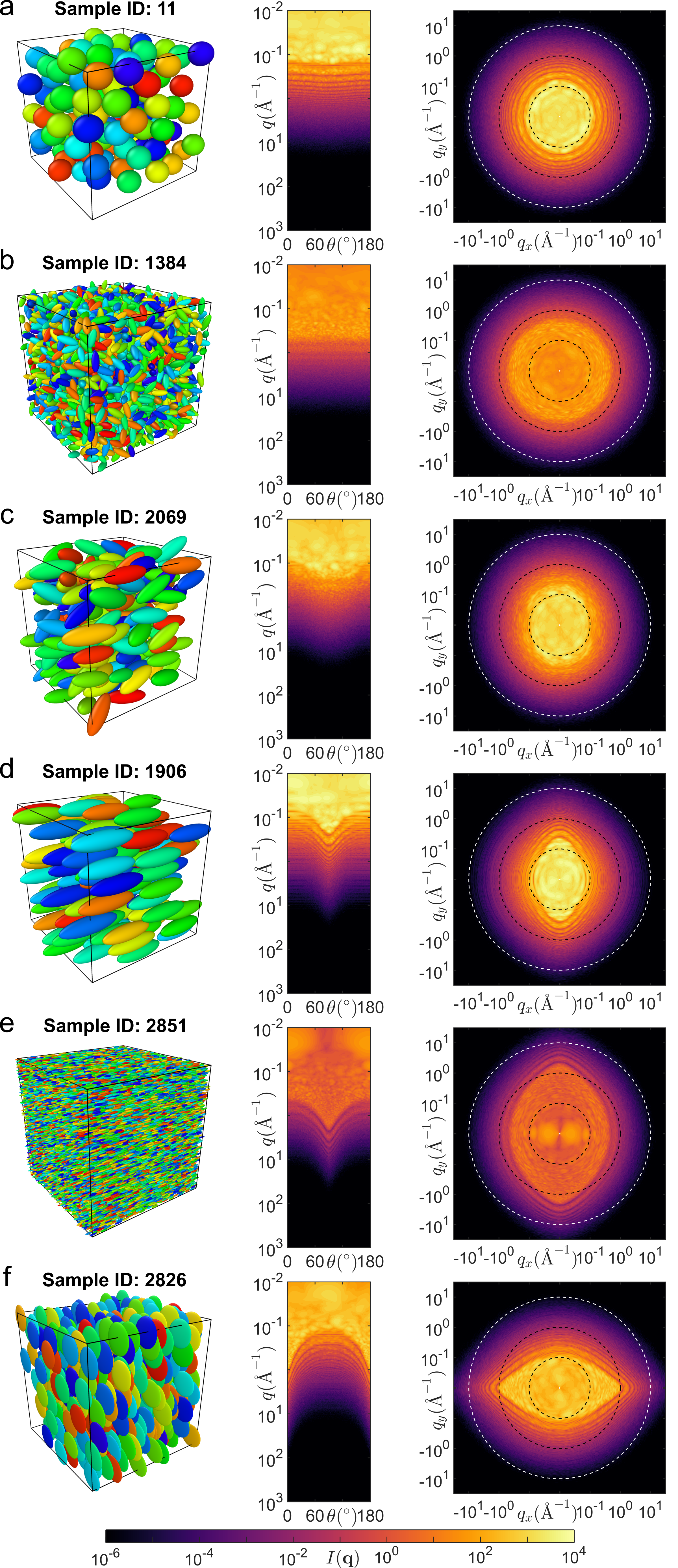}
    \caption{(a-f) From left to right, representative snapshots of a structure with Sample ID denoted on the top, the Cartesian form of their scattering intensity profile $I_\text{comp}(q,\theta)$ with axes $q$ and $\theta$, and the polar form of their scattering intensity profile $I_\text{comp}(q_x,q_y)$ with axes $q_x$ and $q_y$. Here, $q_x$ and $q_y$ are components of the scattering vector $\mathbf{q}$ that are (reciprocally) aligned with the laboratory frame axes $x$ and $y$, respectively. The polar form of the scattering intensity profile maintains the logarithmic scaling of the scattering vector magnitudes, and as a result the center of the profile is not at $q=0$ but truncated to $q=10^{-2}$.}
    \label{fig:scattering}
\end{figure}

In all previous implementations of the CREASE method, we used the pairwise Debye scattering equation to calculate the scattering intensity contribution of $N$ particles with known form factors $f_{1,2,...,N}(q)$ as follows:
\begin{equation}
    I_\text{comp}(q) = \frac{1}{V} \sum_{n=1}^{N}\sum_{m=1}^{N} f_{n}(q) f_{m}(q) \frac{\sin(q r_{nm})}{q r_{nm}}
\end{equation}
The above equation is only applicable for isotropic arrangement of particles, and is obtained by integration over all possible orientations of the scattering vector $\mathbf{q}$, which is equivalent to azimuthal averaging performed on the experimental 2D scattering profiles. Notably, this equation has a double-summation term which necessitates pairwise consideration of particles and their contributions to the scattering profiles, and has the effect of making the scattering calculations computationally intensive and harder to parallelize. Together with the time required to generate structures, the additional time needed to perform scattering calculations for that structure, makes them unfit for use directly within the CREASE workflow. This motivated the need for surrogate ML models that are time-efficient in the prediction of $I_\text{comp}(q)$ (as described in step (3) sub-section).

For CREASE-2D implementation, we compute the 2D scattering intensity $I_\text{comp}(\mathbf{q})$  from the scattering amplitude $A_\text{comp}(\mathbf{q})$ as $I_\text{comp}(\mathbf{q})={1}/{V} \left|A_\text{comp}(\mathbf{q})\right|^2$. Here, the $A_\text{comp}(\mathbf{q})$ is the complex Fourier transform of the fluctuation in the scattering length density $\Delta\rho_{n}$\cite{glatter1982small,guinier1955small,brisard2013small}, and is expressed as:

\begin{equation}
    A_\text{comp}(\mathbf{q}) =  \sum_{n=1}^{N} \Delta\rho_{n} v_{n} f_{n}(\mathbf{q})  \exp\left(-i~\mathbf{q} \cdot \mathbf{r}_{n}\right)
    \label{eqn2}
\end{equation}
The above expression only has a single summation term, which significantly reduces the computational complexity of the scattering calculation from $O(N^2)$ to $O(N)$, and has enabled the computation to be parallelized; this was also noted by Brisard et al. as the `simple sums' computation.\cite{brisard2013small} For our model system, due to the simplicity of the ellipsoidal shape, computation of equation \ref{eqn2}, is further simplified with the anisotropic form factor $f_{n}(\mathbf{q})$ of a spheroid, which can also be obtained from Pedersen's tabulation of analytical form factors.\cite{pedersen1997analysis} $f_{n}(\mathbf{q})$ is provided as:

\begin{equation}
    f_{n}(\mathbf{q}) \equiv f_{n}(q,\theta)=  \frac{j_{1}(q~r_n(\theta))}{q~r_n(\theta)}.
    \label{eqn3}
\end{equation}

In the above equation $j_{1}(\cdot)$ is the first spherical Bessel function, and $r_n(\theta)$ is an effective radius (of particle $n$) that depends on the direction ($\theta$) of the scattering vector $\mathbf{q}$. A more detailed expression for the analytical form factor can be found in the \textbf{Supporting Information Section S2}. We note that for shapes of particles that are complex, without easily available analytical forms of shape, one can calculate the entire 2D scattering profile by placing point scatterers in the box and using sufficient number of point scatterers to resolve the particle shapes and particle-particle spatial arrangements. We are currently finalizing a computational efficient, GPU-based code, to calculate 2D scattering profiles for any shape of the particle using this scatterer approach; we will share that as open-source code on https://github.com/arthiayaraman-lab . 

As the structure is contained in the shape of a cubical box of length $L$, the scattering calculations can be heavily dominated by the form factor of the cubical box, referred to as the `finite size effects' in literature\cite{brisard2013small}. These finite size effects greatly obscure the 2D scattering profile of the structure, and makes it hard to interpret their variation purely due to the structural features. By accounting for the volume fraction of each particle, the form factor of the box can be subtracted as a correction to the scattering profile of the structure. We have adapted the correction scheme described by Brisard et al.\cite{brisard2013small} to remove these finite size effects as discussed in the \textbf{Supporting Information Section S2}. Some simplifications like considering the full shape of the particles at the boundaries can be made and work well as long as the cubic box size is much larger than the size of the particles, which in our case is below $10\% L$. 

\textbf{Figure~\ref{fig:scattering}a-f} provides some representative examples of scattering profile variations computed using equation~\ref{eqn3} after applying the finite size effects correction. In each panel a structure denoted by their Sample ID is shown together with two representations of the computed scattering profiles, which are obtained as the color-coded intensity plots for each $q$ and $\theta$ value. The left scattering plots in Figure~\ref{fig:scattering} are referred to as the Cartesian scattering intensity plots - $I_\text{comp}(q,\theta)$, with axes $q$ and $\theta$. The right scattering plots show the polar scattering intensity $I_\text{comp}(q_x,q_y)$, with axes $q_x$ and $q_y$, two components of the scattering vector $\mathbf{q}$. The polar form is easily recognizable to the soft materials experts, and is the typical representation of 2D scattering profiles directly produced from SAXS/SANS measurements. Both the ``Cartesian'' and ``polar'' forms of 2D scattering profiles are numerically equivalent and only differ in their visual representation. We have included the polar scattering intensity plots as a reference to the reader for easier comparison to relevant experimental scattering profiles. However, for the further analysis, the Cartesian representation provides a convenient and straightforward representation of the numerical data, since only half of the complete plot $\theta=0^\circ$ to $\theta=180^\circ$ needs to be represented due to the inversion symmetry with the other half of the profile, i.e. $I(\mathbf{q})=I(\mathbf{-q})$. As demonstrated further, the Cartesian representation can be easily serialized to obtain the complete training and testing data in a tabular form that is convenient for training the surrogate ML model as described in the next sub-section.

The structures chosen in \textbf{Figure~\ref{fig:scattering}} are used to demonstrate how structural variations can influence the scattering profile. For example, \textbf{Figure~\ref{fig:scattering}a} and \textbf{\ref{fig:scattering}b} each demonstrate an isotropic scattering profile, while having different shapes of the individual particles; more spherical in the former and disordered (low $\kappa$) prolate-spheroidal in the latter. \textbf{Figure~\ref{fig:scattering}c} shows weakly aligned structure (intermediate $\kappa$), while \textbf{Figure~\ref{fig:scattering}d-f} show highly aligned structures (high $\kappa$). Another distinguishing effect is the change in the intensity at low $q$ for \textbf{Figure~\ref{fig:scattering}d} and \textbf{Figure~\ref{fig:scattering}e}; this is due to the drastic change in average size of the particles.

\subsection*{Step (3): Training the Machine Learning (ML) Model to Link Structural Features to Computed Scattering Profile}

With a streamlined implementation of steps (1) and (2), the dataset of 3000 3D structures and their corresponding 2D scattering profiles is ready for the ML model training and testing (or validation). 80\% of the data (2400 structures) are used for training the ML model and remaining 20\% (600 structures) are used for validation of the ML model's performance.

\begin{figure*}[t!]
    \centering
    \includegraphics[width=1\textwidth]{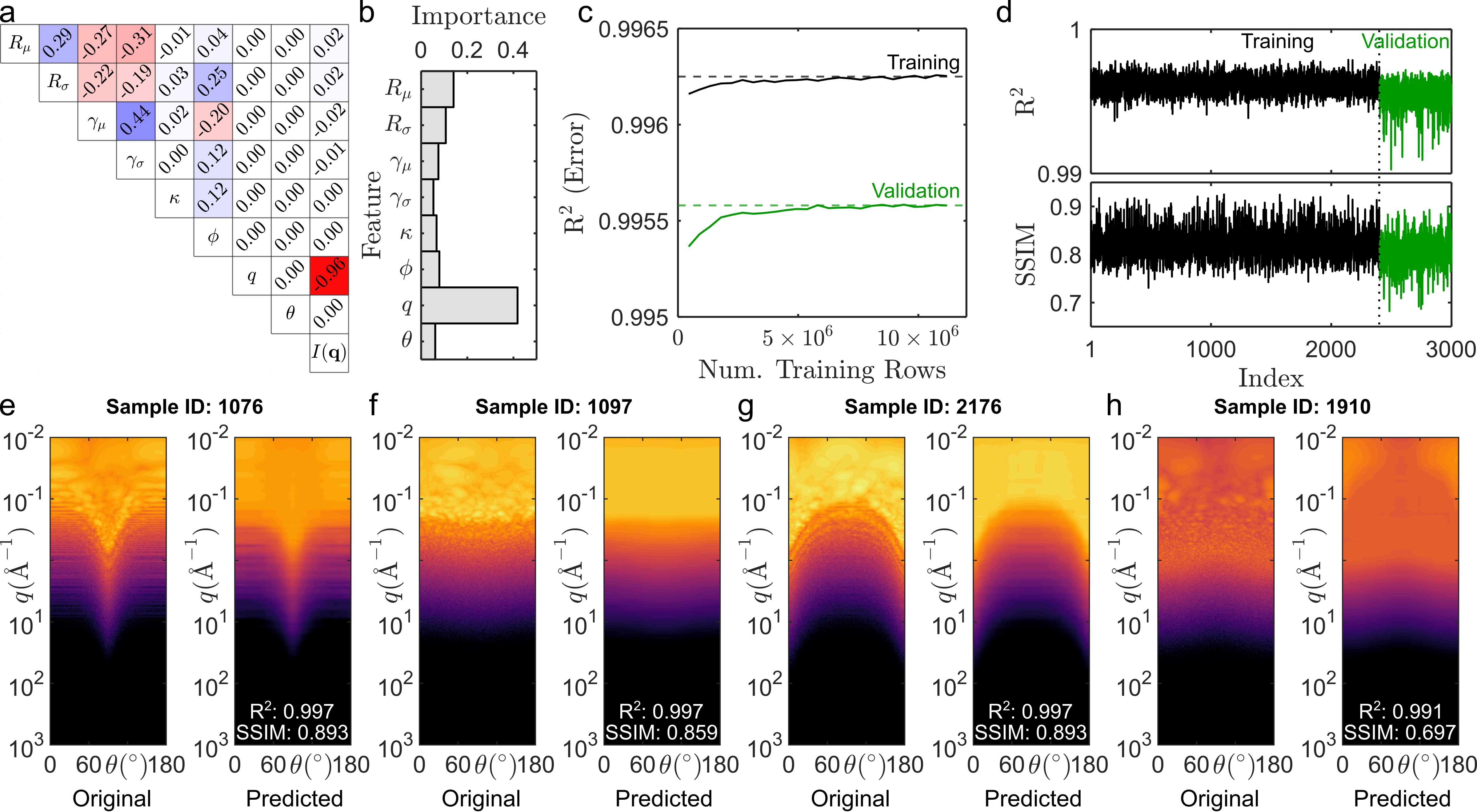}
    \caption{(a) Correlation matrix for all 6 structural features along with $q$, $\theta$ and $I(\mathbf{q})$ values, that together form the dataset for training and validation. (b) The importance histogram for each feature evaluated by the model after training. (c) Learning curve during training of the surrogate ML model, where R\textsuperscript{2} error of the training (black) and validation (green) data entries is plotted against the number of data entries. (d) Performance of the ML model using the R\textsuperscript{2} and the Structural Similarity Index Measure (SSIM) scores for all 3000 samples in the dataset. We note that the index in the x-axis for this plot runs from 1 to 3000 but it is different from Sample IDs; the index distinguishes the randomly selected 2400 samples used for training and 600 samples for the validation. (e-h) Original and predicted scattering profiles for a selected few samples from the validation dataset, each marked with their R\textsuperscript{2} and SSIM scores.}
    \label{fig:MLperformance}
\end{figure*}

In our efforts to apply an appropriate ML method that predicts the 2D scattering profile from a given set of structural features, we identify the need to use a supervised ML approach where continuous-valued quantities can be predicted from a small set of other continuous parameters. Traditionally, both deep learning (DL) and ensemble learning approaches have been successfully applied to achieve these tasks.

With the many DL approaches, one can create a generative model that is conditionally trained on all the structural features. However, successful training of generative models requires a lot more data than provided in our dataset, roughly estimated to be well above 10,000 - 100,000 images for model training alone.\cite{elasri2022image,shrestha2023conditional} On the other hand, ensemble learning methods combine the prediction of multiple standalone models, to create an overall `ensemble' predictive model that is more accurate than the individual predictions from the standalone models. Many ensemble learning approaches can be easily implemented using decision trees which are simpler to work with than neural networks, and have been shown to perform exceptionally well, outperforming neural networks\cite{mcelfresh2023neural} for tabular data, as is also the case for $I_\text{comp}(q,\theta)$.  Motivated by these advantages, we use a decision tree-based ML model to predict the value of $I_\text{comp}(q,\theta)$ for each of the 6 structural features and the given $q$ and $\theta$ values.

In the realm of decision tree-based ML models, especially when dealing with tabular data, boosting ML techniques have gained popularity.\cite{song2020steel,choi2019data,gong2022xgboost,zhang2019application} This is because boosting ML techniques combine groups of weak predictive learners sequentially and correct previous models' training loss to form a strong ensemble model. Here we choose XGBoost algorithm,\cite{xgboost_paper} which stands for e\textbf{X}treme \textbf{G}radient \textbf{Boost}ing, to be the surrogate ML model in CREASE-2D, due to its exceptional performance and lower scope of over-fitting. 

XGBoost is a generalized algorithm that can be implemented to solve a wide range of problems. During training, XGBoost assigns weights to all features it trains on, referred to as feature importance and accordingly adjusts the construction of decision trees.  XGBoost also offers a wide range of hyper-parameters that can be fine-tuned to a diverse set of training data. In our work, we utilize these advantages of XGBoost to train the surrogate ML model that outputs a 2D scattering intensity for the input of structural features and $(q,\theta)$ values.

To use XGBoost algorithm, the training dataset is reformatted into a table, where each row contains all the 6 structural features as fields, combined with a serialized representation of the scattering profiles. Thus, each training dataset row reads as: $R_\mu$, $R_\sigma$, $\gamma_\mu$, $\gamma_\sigma$, $\kappa$, $\phi$, $q$, $\theta$, and $I(q,\theta)$. During serialization of the dataset, the resolution of the scattering profile can have a dominant effect on the efficiency of training. This is because  a higher resolution will result in better quality of the data, but also increases the computational overhead and memory requirements during training. The 2D scattering profiles calculated in step (2) are generated over a ($q$, $\theta$) grid of $501 \times 181 = 90681$ data points, which amounts to over 200 million points for all 2400 samples in the training dataset. In principle one could use all these points to train the surrogate ML model, if the user has outstanding computational resources with limitless memory. For users with modest computational resources (including cost-effective subscriptions to Google-Colab) sub-sampling of the data is deemed necessary. We therefore adopted a grid-based sub-sampling approach where we uniformly sample every 4\textsuperscript{th} $q$ value and every 5\textsuperscript{th} $\theta$ value to obtain a ($q$, $\theta$) grid of $127 \times 37 = 4662$ data points. This results in around 11 million tabular entries for the 2400 samples that can be handled reasonably well by the ML model.

To tune the architecture of the decision trees in the XGBoost model, Bayesian search optimization\cite{thebelt2022tree} with cross validation is performed over a large range of hyper-parameters to identify their best configuration that provides reliable accuracy in the predicted 2D scattering profiles. More details about configurations of Bayesian optimization is provided in the \textbf{Supporting Information Section S3}. As an example, after this optimization, we find that the predicted intensity values are the most reliable when for each decision tree and for each node of a decision tree only $90\%$ and $80\%$ of the structural features are randomly sampled, respectively. Other hyper-parameters which determine the learning rate, step-size, maximum depth of the decision tree, etc. are also optimized and  described in more detail in the \textbf{Supporting Information Section S3} along with their optimum value that are used to train the ML model. Careful tuning of these hyper-parameters is essential for achieving optimal model performance and avoiding over-fitting on the given dataset. Bayesian optimization of the hyper-parameters takes just over an hour to optimize, when using the V100 GPUs with 51 GB RAM as provided by our Google Colab Pro subscription. Once the tuned hyper parameters are obtained, the XGBoost model is trained on CPUs within 10 minutes.


To understand the data, we present the correlation matrix in \textbf{Figure~\ref{fig:MLperformance}a} and to understand how the ML model interprets the data after training we present the histogram that measures the feature importance in \textbf{Figure~\ref{fig:MLperformance}b}. The correlation matrix indicates weak correlations between the means and standard deviations of $R$ and $\gamma$, possibly due to the way these values are sampled, as indicated in Step (1). Some correlations are also observed for $\phi$ and all remaining structural features; as noted above in the CASGAP structure generation, the volume fraction $\phi$ value is not directly varied during structure generation and is only evaluated after the structure is generated. The strongest (inverse) correlation is observed between the scattering intensity $I(q,\theta)$ and the magnitude of the wavevector $q$; this is expected as the scattering intensity values display a drastic dependence on the $q$ values. Consequently, after training, the ML model assigns the highest importance to $q$, as shown in \textbf{Figure~\ref{fig:MLperformance}b}. \textbf{Figure~\ref{fig:MLperformance}c} shows the learning curve where the performance is measured using the R\textsuperscript{2} error, which is a normalized version of the mean squared error (MSE), and is plotted against the number of data entries that the model has already used for training. Both the training and the validation errors are found to converge quickly to beyond 99.5\% indicating that the surrogate ML model does not over fit the training data.

In \textbf{Figure~\ref{fig:MLperformance}d} we evaluate the performance of the surrogate ML model using two metrics for all 3000 samples, where we have assigned an index (different from Sample ID) to separate the training samples from the validation (or test) samples. The first metric is the R\textsuperscript{2} error evaluated in a similar way as done during ML model training. These R\textsuperscript{2}  scores provide information about the prediction accuracy of the ML model at each $q$ and $\theta$ value on an individual basis (i.e., without necessarily considering the local context). We find that the R\textsuperscript{2} values converge to 0.995 and do not differ much for training vs. validation samples, indicating excellent prediction accuracy of the ML model. However, to also evaluate the performance of the ML model to output the entire 2D scattering profile we need another metric that takes into account the performance of the model at all values in the local vicinity of a $q$ and $\theta$. For this reason, we choose the structural similarity index (SSIM) scores which infers the structural differences between the two scattering profiles, by using image-based characteristics like luminescence, contrast, and pattern; these quantities are derived from the mean, variance, and covariance information of the local pixel data. An SSIM score near 1 indicates a good performance of the ML model in predicting the entire scattering profile for a given set of structural features. In \textbf{Figure~\ref{fig:MLperformance}d} the SSIM scores converge to above $\sim0.8$ indicating a reliable prediction accuracy of the ML model.

A visual comparison between the original and the ML predicted scattering profiles is also shown in  \textbf{Figure~\ref{fig:MLperformance}(e-h)}, along with their R\textsuperscript{2} and SSIM scores. We note that among all the Sample IDs,  Sample 1910 shown in this figure has the least SSIM score. A more detailed comparison between the original and predicted profiles is provided in the \textbf{Supporting Information Section S4}, by overlaying their 1D scattering profiles at a few selected $\theta$ values to further demonstrate the similarities in the two profiles. These results demonstrate that the trained surrogate ML model performs reasonably well. It is important to note that the quality of the surrogate model training will impact how well CREASE-2D performs. We encourage users of CREASE-2D to invest the appropriate time for the ML model training, and to ensure that poor training and testing do not manifest as poor analyses from CREASE-2D.

\subsection*{Step (4): Optimization within Genetic Algorithm (GA) in CREASE-2D}

\begin{figure*}[t!]
    \centering
    \includegraphics[width=\textwidth]{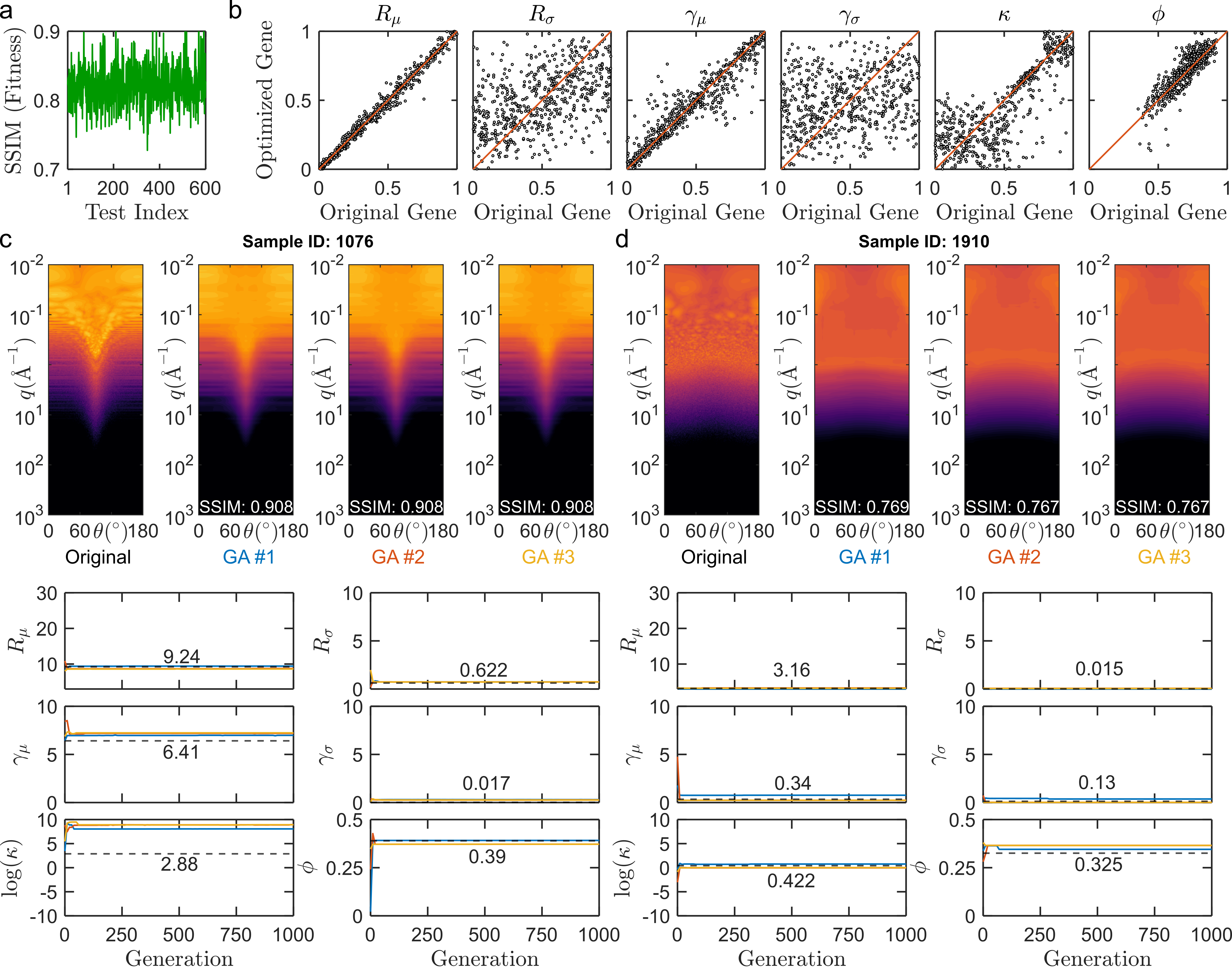}
    \caption{Performance of Genetic Algorithm (GA)  in CREASE-2D. (a) Structural Similarity Index or SSIM scores for all 600 samples input to CREASE-2D; SSIM quantifies the similarity between the GA-optimized or `best' $I_\text{comp}(q, \theta)$ at the end of the GA loop with the input $I_\text{exp}(q,\theta)$. (b) The comparison of GA-optimized values of the normalized ``gene'' or structural features and the original value of the structural feature, normalized to represent a target gene value for all 600 samples tested with CREASE-2D. (c-d) Two selected samples - Sample IDs 1910 and 1076- out of the 600 samples tested with CREASE-2D. We show visual comparison of the input scattering profile and outputs from three independent GA runs and plot their corresponding evolution of structural feature predictions during each GA run for Sample IDs 1076 and 1910. The solid colored curves in the plots in (c) and (d) are the three GA runs and the black dashed curve is the value of the structural feature corresponding to the original scattering, with the exact value of that structural feature denoted in text. }
    \label{fig:GAperformance}
\end{figure*}

The final step in CREASE-2D implementation is to put together the predictive capacity and the speed of the surrogate ML model within the genetic algorithm (GA) optimization loop. We refer the reader to previous CREASE publications\cite{beltran2019computational,heil2023computational,heil2022computational,wessels2021computational,wessels2021machine,wu2022machine,ye2021computational,heil2021computational,crease_ga} for detailed implementations of the GA optimization loop in the successful execution of the CREASE (1D) method. In the current implementation, one major distinction is the use of a continuous parameter GA in contrast to the binary GA used in the previous work. The continuous parameter GA is better suited for evolving ``genes'' that represent continuous parameters, and has a more straightforward interpretation of the crossover and mutation operations.\cite{haupt2004practical} As noted before, the 6 structural features ($R_\mu$, $R_\sigma$, $\gamma_\mu$, $\gamma_\sigma$, $\kappa$, $\phi$) are represented as 6 corresponding ``genes'' and every ``individual" has a unique set of values for these genes in the GA optimization loop . We first normalize values of the genes, using a scheme similar to the one used to obtain their randomized distribution; for more detail see \textbf{Supporting Information Section S5}. The normalization schemes assigns a value between $0 - 1$ as the value of the gene and has a monotonic one-to-one correspondence with the value of the corresponding structural feature.

For every ``individual'' with a unique set of genes, a scattering profile is predicted from the surrogate ML model using the individual's structural features as the input.  All individuals in each generation are then ranked by their ``fitness'' value which is quantified by the structural similarity index (SSIM) of the individual's computed scattering profile with respect to the experimental input scattering profile. The objective of the GA optimization loop is to improve the fitness of an individual; in other words,  improve the SSIM score of its computed scattering profile $I_\text{comp}(q,\theta)$ as compared to $I_\text{exp}(q,\theta)$.

The other important considerations in the implementation of the GA optimization loop is the choice of the number of individuals to sample in each generation (i.e., the population size) as well as the selection procedures for determining individuals that move to the next generation.  In our implementation, we use a fixed population of 100 individuals per generation that are always ranked according to their fitness. In each generation, the top 30 individuals with the highest fitness are selected. These 30 individuals serve as parents who are randomly paired to form 70 children using a single-point cross-over method. Subsequently, the 30 parents and 70 children together form the 100 individuals for the next generation. For these 100 individuals, the next set of operations are related to mutation. The top two elite individuals' gene values are kept unchanged as they progress to the next generation. The remaining 98 individuals undergo adaptive mutation, where the mutation probability and step size is varied based on the L2 distance (or the squared Euclidean distance) of the individual from the mean value of all individuals. Adaptive mutation is usually recommended to prevent the GA from converging too quickly to a local minimum, and to have sufficient diversity in the genes and individuals in the population.\cite{marsili2000adaptive} With this next generation of 100 individuals, the GA optimization loop is then continued. As the number of generations increases, the fitness of a generation should converge, and upon convergence the GA loop can be stopped. 

\subsection*{Performance of CREASE-2D}

To evaluate the performance of the CREASE-2D method, we use all the 600 test samples out of the 3000 test samples (the reader will recall that out of the dataset of 3000 samples, 2400 samples are used to train the surrogate model) and run GA five separate times for each sample. We run five GA separate runs for each sample to check how different the output structural features from CREASE-2D are for each sample's input scattering profile; this allows us to understand degeneracy in optimized GA solutions. In \textbf{Figure~\ref{fig:GAperformance}\textbf{a}} the fitness measured in the form the SSIM scores for all 600 samples tested in CREASE-2D are all found to in the 0.7 to 0.9 range. We note that this SSIM score range is similar to the performance of the surrogate ML model indicating a reliable match between the input and output scattering profiles. For each sample, the standard deviations in the SSIM scores from its five GA replicates  are small (within 1\% of the value shown) and thus, not shown in the plot for clarity. We note that the CREASE-2D method performs only as well as the surrogate ML model and it should not be expected to outperform the prediction accuracy of the surrogate ML model.

In \textbf{Figures~\ref{fig:GAperformance}b}, we compare how well CREASE-2D predicts each structural parameter value for all 600 test sample whose structural features we (but not CREASE-2D) know \textit{a priori.} To make an effective visual comparison, we use the gene values directly instead of the structural features in these plots. For some structural features, especially  $\phi$, $R_\mu$ and $\gamma_\mu$ (i.e., volume fraction and means of particle size and shape distributions) the accuracy of prediction is high, as indicated by the clustering of points close to red line with unit slope. For  $R_\sigma$ and $\gamma_\sigma$  (which measure of the dispersity in particle size and shape) the prediction accuracy is low, despite having a high SSIM score. This indicates that precise values of extent of dispersity in the particle size and shape have a minimal impact on the variation of the scattering intensity; this is in line with observations in experiments that presence of dispersity broadens peaks of scattering profile but does not alter the shape of the profile with the value of dispersity. As a result, for $R_\sigma$ and $\gamma_\sigma$,  CREASE-2D method is dealing with larger degeneracy in solutions. For  $\kappa$,  that quantifies orientational order, the accuracy is high only for samples that have high values of $\kappa$, and the accuracy is low for samples with lower $\kappa$ values (i.e., low orientational order). As one would expect, at low values of $\kappa$ which represent isotropic ordering of anisotropic particles, the precise numerical value of $\kappa$ value has minimal impact on the scattering intensities.

To further demonstrate the performance of CREASE-2D method for the four representative samples (same as from \textbf{Figure~\ref{fig:MLperformance}}), in \textbf{Figure~\ref{fig:GAperformance}c} and \textbf{\ref{fig:GAperformance}d} the results for Sample IDs 1076 and 1910 are presented and the results for Sample IDs 1097 and 2176 are provided in \textbf{Supporting Information Section S5}. The evolution of their fitness values is also presented in \textbf{Supporting Information Section S5}. In \textbf{Figure~\ref{fig:GAperformance}} we show the predicted scattering profiles for the best outputs from 3 out of the 5 GA runs per system along with the evolution of structural features over 1000 generations from those 3 GA runs. In each run, the GA loop converges closely to the original value of the structural feature in the first few generations, as indicated by the convergence of the curves to the dashed line (the numbers in the plots denote the target structural feature value of that sample). We note that one GA optimization loop with 1000 generations of 100 individuals uses 30-45 minutes in real-time to complete when implemented on a single-(CPU) core laptop/computer with modest hardware.

In conclusion, we have developed a new CREASE-2D method that analyzes 2D scattering profiles as is without any averaging along all or few angles, and outputs relevant structural features like domain size and shape distribution, extent of orientational order in the structure, and packing fraction of the domains in the structure. The development of CREASE-2D relied on the generation of dataset with 3000 samples each with a desired set of structural features and the corresponding 3D structures generated using CASGAP \cite{gupta2023} and corresponding computed 2D scattering profile.  This dataset enabled training of a surrogate XGBoost-based model that outputs 2D scattering profile for a given set of structural features. Using this surrogate ML model within a genetic algorithm (GA) optimization loop, we are able to identify all the structural features (and reconstruct 3D real-space configurations, if needed) that produce a scattering profile that matches the input 2D scattering profile. We believe soft materials researchers who aim to understand how macroscopic properties (e.g., rheology, flow) depend on the structural anisotropy and the hierarchy of structural length scales within the materials will find this CREASE-2D method useful. CREASE-2D enables users to analyze the output of scattering experiments holistically without having to use approximate analytical models to fit to averaged 1D profiles or limiting to analyzing only averaged angular sections of the 2D profiles. 

\section*{Acknowledgements}
We are grateful for financial support from National Science Foundation (NSF) Grant number: 2105744 (for SA and AJ) and MURI AFOSR Grant number MURI-FA 9550-18-1-0142  (for NG and AJ) to complete this method development work. We also acknowledge the use of DARWIN computing system for some of the computing in this work: DARWIN – A Resource for Computational and Data-intensive Research at the University of Delaware and in the Delaware Region, Rudolf Eigenmann, Benjamin E. Bagozzi, Arthi Jayaraman, William Totten, and Cathy H. Wu, University of Delaware, 2021.

\section*{Supplementary Information}
The Supporting Information is available free of charge at website-to-be-added.
Includes additional views of results presented in the main paper as well as specific details about some of the steps in the method development. 

\section*{Code and Data Availability}
Open-source code for CREASE-2D is available on https://github.com/arthijayaraman-lab/
The dataset discussed in the paper are available on https://zenodo.org/records/10534943
The open-source code for 3D structure creation method, CASGAP, is also available on https://github.com/arthijayaraman-lab/

\section*{Author Contributions}
AJ, SA, and NG developed the ideas for this CREASE-2D method. AJ received funding to support this project. NG and SA generated the dataset of structures and scattering profiles. SA developed the surrogate XG-Boost ML model to train on the dataset and tested the model. NG and SA created the CREASE-2D workflow and implemented the genetic algorithm optimization. AJ, NG, and SA wrote and edited the manuscript.  


\bibliographystyle{plain}

\end{document}